\documentclass[10pt,journal,twoside]{IEEEtran}
\usepackage{amssymb}
\usepackage{amsthm}
\usepackage{multirow}
\usepackage{inputenc}
\usepackage{enumerate}
\usepackage{enumitem}
\usepackage{textcomp}
\usepackage{listings}
\usepackage{array}
\usepackage[switch,mathlines]{lineno}
\usepackage{soul}
\usepackage{xfrac}
\usepackage{graphicx}
\usepackage{cite}
\usepackage[cmex10]{amsmath}
\usepackage{xcolor}
\usepackage{colortbl}
\usepackage{cancel}
\usepackage{siunitx}
\usepackage{nomencl}
\usepackage{multirow}
\usepackage{svg}
\usepackage{doi}
\usepackage{hyperref}
\usepackage{amsmath}
\usepackage{subfig}
\usepackage{verbatim}
\usepackage[ruled,norelsize,vlined,linesnumbered]{algorithm2e}
\makeatletter
\newcommand{\removelatexerror}{\let\@latex@error\@gobble}
\makeatother
\newtheorem{myDef}{Definition}

\makeatletter
\def\footnoterule{\kern-3\p@
  \hrule \@width 3.3in \kern 2.6\p@} 
\makeatother

\usepackage{algpseudocode}

\usepackage{stackengine}

\usepackage{algpseudocode}
\usepackage{CJK}
\usepackage[cmex10]{amsmath}
\usepackage{bm}
\usepackage{color}
\usepackage{amsmath,amsthm,amssymb,amsfonts}
\usepackage{flushend}
\usepackage{float}

\newtheorem{lemma}{Lemma}
\newtheorem{remark}{Remark}
\newtheorem{proposition}{Proposition}
\usepackage{arydshln} 
\usepackage{hyperref}
\hypersetup{
     colorlinks   = true,
     linkcolor    = blue,
     citecolor    = blue,
     urlcolor     = blue
}
\makeatletter
\newcommand*{\transpose}{%
  {\mathpalette\@transpose{}}%
}
\newcommand*{\@transpose}[2]{%
  \raisebox{\depth}{$\m@th#1\intercal$}%
}
\makeatother

\usepackage{soul}
\usepackage{tikz}
\usepackage{booktabs}
\usepackage{tabularx}
\usepackage{setspace}

\ifCLASSINFOpdf
\else
\fi
\renewcommand\nomgroup[1]
{
  \ifthenelse{\equal{#1}{A}}{\item[\textit{A. Abbreviations}]}{%
  \ifthenelse{\equal{#1}{S}}{\item[\textit{B. Sets}]}{
  \ifthenelse{\equal{#1}{V}}{\item[\textit{C. Parameters and Variables}]} \\
  }
  }
  }
\begin{document}
\renewcommand{\ttdefault}{cmtt}
\bstctlcite{IEEEexample:BSTcontrol}
\SetKwComment{Comment}{/* }{ */}
\renewcommand\qedsymbol{$\blacksquare$}

\title{Privacy-Preserved Aggregate Thermal Dynamic Model of Buildings}
\author{
{
Zeyin Hou,~\IEEEmembership{Student Member, IEEE},
Shuai Lu,
Yijun Xu,~\IEEEmembership{Senior Member, IEEE},
Haifeng Qiu, \\
Wei Gu,~\IEEEmembership{Senior Member, IEEE},
Zhaoyang Dong,~\IEEEmembership{Fellow, IEEE},
Shixing Ding~\IEEEmembership{}
}

\thanks{The work was supported by the National Natural Science Foundation of China (52207080), in part by the Jiangsu Provincial Key Laboratory of Smart Grid Technology and Equipment, Southeast University, and in part by the Fundamental Research Funds for the Central Universities (\emph{Corresponding author: Shuai Lu}).}
\thanks{Z. Hou, S. Lu, Y. Xu, W. Gu are with the  Electrical Engineering Department, Southeast University, Nanjing, Jiangsu 210096, China, (e-mail: \href{mailto:hzy17301571262@163.com}{hzy17301571262@163.com}; \href{mailto:shuai.lu.seu@outlook.com}{shuai.lu.seu@outlook.com}; \href{mailto:yijunxu@seu.edu.cn}{yijunxu@seu.edu.cn}; \href{mailto:wgu@seu.edu.cn}{wgu@seu.edu.cn}).

H. Qiu and Z.Y. Dong are with the School of Electrical \& Electronics Engineering, Nanyang Technological University, Singapore 639798 (e-mail: \href{mailto:haifeng.qiu@ntu.edu.sg}{haifeng.qiu@ntu.edu.sg}; \href{mailto:zy.dong@ntu.edu.sg}{zy.dong@ntu.edu.sg}).

S. Ding is with the School of Cyber Science and Engineering, Southeast University, Nanjing 210096, China (e-mail: \href{mailto:dingshx616@163.com}{dingshx616@163.com}).}
}

\markboth{S\MakeLowercase{ubmitted to} IEEE Transactions on Smart Grid}
{Hou \MakeLowercase{\textit{\textit{et al.}}}: Privacy-Preserved Aggregate Thermal Dynamic Model of Buildings}
\maketitle

\begin{abstract}
The thermal inertia of buildings brings considerable flexibility to the heating and cooling load, which is known to be a promising demand response resource. The aggregate model that can describe the thermal dynamics of the building cluster is an important interference for energy systems to exploit its intrinsic thermal inertia. However, the private information of users, such as the indoor temperature and heating/cooling power, needs to be collected in the parameter estimation procedure to obtain the aggregate model, causing severe privacy concerns.
In light of this, we propose a novel privacy-preserved parameter estimation approach to infer the aggregate model for the thermal dynamics of the building cluster for the first time. 
Using it,  the parameters of the aggregate thermal dynamic model (ATDM) can be obtained by the load aggregator without accessing the individual's privacy information. 
More specifically, this method not only exploits the block coordinate descent (BCD) method to resolve its non-convexity in the estimation but investigates the transformation-based encryption (TE) associated with its secure aggregation protocol (SAP) techniques to realize privacy-preserved computation.
Its capability of preserving privacy is also theoretically proven. 
Finally, simulation results using real-world data demonstrate the accuracy and privacy-preserved performance of our proposed method.
\end{abstract}


\IEEEpeerreviewmaketitle

\makenomenclature
\nomenclature[A]{ATDM}{Aggregate thermal dynamic model}
\nomenclature[A]{BLA}{Building load aggregator}
\nomenclature[A]{TE}{Transformation-based encryption}
\nomenclature[A]{SAP}{Secure aggregation protocol}
\nomenclature[A]{MQS}{Multivariate quadratic system}
\nomenclature[A]{BCD}{Block coordinate descent}
\nomenclature[A]{LSR}{Least squares regression}
\nomenclature[S]{$\mathbf{M}$}{Index set of model order}
\nomenclature[S]{$\mathbf{T}$}{Index set of time periods}
\nomenclature[S]{$\mathbf{K}$}{Index set of building zones}
\nomenclature[S]{$\mathbf{L}$}{Index set of iterations}
\nomenclature[V]{\(M, m\)}{Model order and its index}
\nomenclature[V]{\(K, i\)}{The total number of zones in the building cluster and its index}
\nomenclature[V]{\(L,l\)}{The total number of iterations and its index}
\nomenclature[V]{\(T,t\)}{The total number of period and its index}
\nomenclature[V]{\(\alpha_{(\cdot)}, \beta_{(\cdot)}, \gamma_{(\cdot)}, \theta_{(\cdot)}\)}{Parameters of the ATDM}
\nomenclature[V]{\(\tau_{in,z}^{i,t}\)}{Indoor temperature of building zone $i$ at period $t$}
\nomenclature[V]{\(\tau_{out}^t\)}{Outdoor temperature at period $t$}
\nomenclature[V]{\(h_{load,z}^{i,t}\)}{Heating/cooling power of building zone $i$ at period $t$}
\nomenclature[V]{\(h_{rad}^t\)}{Solar radiation power at period $t$}
\nomenclature[V]{\(\tau_{occ,bc}^t\)}{Impact of occupants' activities at period $t$}
\nomenclature[V]{\(\tilde{\tau}_{in,bc}^t\)}{Aggregate indoor temperature of building cluster at period $t$}
\nomenclature[V]{\(\xi_{bc}^i\)}{Aggregation coefficient of building zone $i$}
\nomenclature[V]{\(\lambda\)}{Penalty factor}
\nomenclature[V]{\(\varepsilon\)}{Random error}
\nomenclature[V]{\((\cdot)^{[i]}\)}{The $i$-th column of matrix $(\cdot)$}
\nomenclature[V]{\({W}\)}{Random matrix}
\nomenclature[V]{\(\tilde{(\cdot)}\)}{The masked version of ($\cdot$) after implementing the SAP method}
\printnomenclature

\section{Introduction}
\IEEEPARstart{T}{he} heating and cooling demands of buildings take up a large proportion of energy consumption \cite{bourdeau2019modeling}, playing an important role in the energy system. The inherent storage capacity of buildings brings considerable flexibility to the heating and cooling demands \cite{pan2017feasible}, which is vital for the operation and control of the energy system \cite{8063955, zhao2015dynamic}. 
Therefore, a thermal dynamic model that can accurately describe the relationship between control and state variables of buildings is a must \cite{fontenot2019modeling}.

Apparently, establishing an accurate model for each building facilitates the refined control of heating and cooling power. However, due to prohibitive computation complexity and heavy communication burden, it is impractical for the energy system to interact with numerous buildings in a peer-to-peer way directly. In practice, the buildings in one area are usually clustered to access the energy system in an aggregated manner. Hence, an aggregate model that can describe the overall thermal dynamics of the building cluster is an important interface with the energy system \cite{song2018state}.  

In existing research, three different approaches have been proposed to obtain the aggregate model of the building cluster, including the physics-based method, the data-driven method \cite{lu2021data}, \cite{zhan2021data}, and the hybrid approach \cite{li2021modelling}. The physics-based method builds thermal behavior modeling based on solving equations of energy conversion law \cite{foucquier2013state}. It is overall more accurate than other models but has high model complexity. The data-driven method typically involves linear regressions \cite{lam1997regression}, artificial neural network \cite{kalogirou2000applications}, etc. Although suitable for the unclear physical model, it requires a large quantity and high quality of data while exhibiting poor interpretability. The hybrid method couples the physics-based method and the data-driven method, overcoming their respective drawbacks to a certain extent, like the aggregate thermal dynamic model (ATDM) \cite{lu2021data} and resistance-capacitance model designed for buildings \cite{muthalib2016physically}. Therefore, the hybrid method is gaining increasing popularity in aggregate modeling.

For the latter two data-related methods, with the aggregator, e.g., the building load aggregator (BLA), assumed to calculate the aggregate model of the building cluster using the real indoor temperature and heating/cooling power information of each building (zone) as most existing literature suggested \cite{lu2021data}, \cite{guo2021aggregation}, the users' private information faces a severe risk of disclosure. 
More seriously, once this information is disclosed, the BLA may manipulate the heat/cooling energy price to gain more benefits at the expense of users. Therefore, privacy concern is an unavoidable problem in the aggregate modeling of the building cluster.  It is urgent to develop a privacy-preserved computation approach for the aggregate modeling of the building cluster. Therefore, this is chosen as the focus of our article. 

Here, for the privacy-preserved computing problem, numerous methods have been exploit. Typical examples involve the alternating direction method of multipliers (ADMM), differential privacy (DP) method, homomorphic encryption (HE) method, and transformation-based encryption (TE) method \cite{gonccalves2021critical}. Among them, the ADMM is a decomposition-based algorithm that could protect privacy to some extent but requires the exchange of intermediate information during iterations \cite{boyd2011distributed, ling2013decentralized}.  It leads to serious privacy disclosure in many situations \cite{zhang2018admm},  which have been revealed in the source location problem \cite{alanwar2017proloc}, the agreement problem \cite{mo2016privacy} and the regression problem \cite{mateos2010distributed}. The DP technique protects privacy by adding customized noise to the private data. For example, Dvorkin \emph{et al.}\cite{dvorkin2020differentially} initiate the DP-based optimal power flow (OPF) problem. Wang \emph{et al.} \cite{wang2023differentially} propose the DP-based consensus + innovations method to realize the distributed parameter estimation while preserving the private information of agents. Although this DP method protects privacy to some extent, it essentially alternates the original optimization problem, resulting in an inevitable loss of the optimality \cite{mak2019privacy}. Alternatively, the HE allows computations to be performed on the encrypted data. For instance, Wu \emph{et al.}  propose a privacy-preserved distributed OPF algorithm based on partially homomorphic encryption \cite{9444341}. Chen \emph{et al.}\cite{chen2018privacy} advocate the HE fitting the coefficients of ridge linear regression. Despite the robustness of HE to privacy attacks, the extremely high computational complexity hinders its practical application \cite{gonccalves2021critical, tran2019privacy}. Besides, some researchers have proposed the TE technique, which can transform the original model into its equivalent data-masked one through masking using random matrices. Moreover, Tian \emph{et al.} \cite{tian2021privacy} propose the TE-based method to solve the privacy-preserved mixed-integer quadratic optimization problem for energy management. Karaca  \emph{et al.} \cite{karaca2023masking} offer the TE approach to tackle data privacy problems of both primal and dual variables in a collaborative network revenue management problem. Jia \emph{et al.} \cite{jia2022chance} use the TE technique to solve chance-constrained optimal power flow with private information among agents. 

Although the above-mentioned privacy-preserved computation methods have made many advances in the energy optimization field, developing a privacy-preserved aggregate modeling method for buildings has barely received attention. Particularly,  the parameter estimation of the aggregate model of buildings is a nonconvex optimization problem \cite{lu2021data}. It is difficult to directly apply these privacy-preserved methods that mainly focus on convex optimization problems \cite{karaca2023masking}, \cite{jia2022chance} that is unsuitable for the aggregate modeling of buildings.

Thus, we are motivated to design a privacy-preserved computation method for the aggregate thermal dynamic model (ATDM) of the building cluster.
Based on this method, the BLA can estimate the parameters of the ATDM without knowing the real indoor temperature and heating/cooling power of users. We also prove that, under very mild assumptions, the BLA cannot infer the private information of the users. 

The main contributions are summarized as follows.

\begin {enumerate}
\item {We propose a privacy-preserved algorithm for the aggregate modeling of the building cluster. This method can estimate the parameters of the ATDM without exposing the privacy information of users,  such as the indoor temperature and heating/cooling power.}
\item{As far as we know, we initiate the study of the nonconvex parameter estimation problem with privacy-preservation. We exploit the block coordinate descent (BCD) method to handle the non-convexity and the TE and security aggregation protocol (SAP) methods to mask the privacy information of users.}
\item {We provide a detailed privacy analysis for the proposed method. We reveal that the essence of the privacy inferences of this method is solving multivariate quadratic systems (MQS), which is an NP-hard problem having multiple solutions. This theoretically ensures the privacy preservation performance of the proposed method.}
\end {enumerate}

The rest of this paper is organized as follows. Section II introduces the ATDM for the building cluster and the corresponding parameter estimation method. Section III proposes the privacy-preserved parameter estimate algorithm. Section IV carries out the privacy analysis for the proposed algorithm. Section V gives the numerical simulation results, and Section VI concludes this paper. 

\section{Problem Description}
Usually, there are two different load control modes for buildings, i.e., the direct and indirect load control \cite{lu2021data}, depending on whether the energy system controls the heating/cooling power of each building zone (direct) or the total heating/cooling power of the building (indirect). In this paper, we formulate the proposed privacy-preserved ATDM based on direct load control, which can be extended to the indirect one. For the sake of generality, we use "agent" below to refer to the building zone or the building. In the following, we introduce the ATDM of the building cluster that was initially proposed in \cite{lu2021data}. Then, we present the parameter estimation model of the ATDM without privacy preservation.

\subsection{Aggregate Thermal Dynamic Model}
 
\begin{figure}[t]
 \centering
 \footnotesize
 \includegraphics[width=1\linewidth]{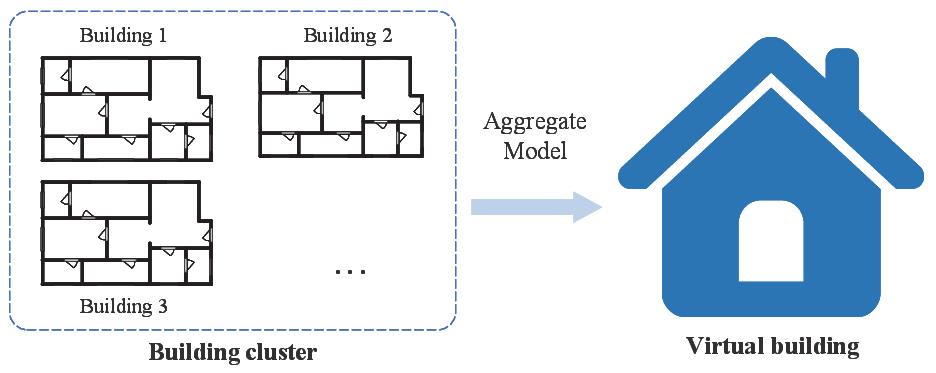}
 \caption{Concept of the Aggregation of the building cluster.}
 \label{fig_aggregation_of_building_cluster_into_virtual_building}
 \end{figure}

The aggregate model of the building cluster under direct control is illustrated in Fig. \ref{fig_aggregation_of_building_cluster_into_virtual_building},  which aims to select a proper state variable to represent the overall characteristics, i.e., the aggregate state. First, following traditions \cite{lu2021data}, \cite{benzaama2020data}, \cite{yun2012building}, we introduce the $M$-order linear time-invariant dynamic system to describe the thermal dynamic evolution of the agent as
\begin{equation}
    \label{building zone model}
    \begin{aligned}
        \tau_{in,z}^{t}&=\sum_{m\in \mathbf{M} \backslash \{0\}}{\alpha_z^m{\tau}_{in,z}^{t-m}+\sum_{m\in \mathbf{M}}{\beta_z ^{m}h_{load,z}^{t-m}}} \\
        &+\sum_{m\in \mathbf{M}}{\gamma_z^{m}\tau _{out}^{t-m}}
        +\sum_{m\in \mathbf{M}}{\theta_z^{m}h_{rad}^{t-m}} +\tau _{occ}^{t}, \ \forall t\in \mathbf{T},
    \end{aligned}
\end{equation}
wherein $\mathbf{M}=\left[0,\cdots, M\right]$ and $\mathbf{T} = \left[1, \cdots, T \right]$. Considering that occupant activities are usually periodical, $\tau_{occ}^t$ is set to a time series with a period of $T_{occ}$.

Then, we use the linear aggregation method proposed in \cite{lu2021data} to derive the aggregate state. From the perspective of energy, the aggregate state should be able to quantify the energy storage in the buildings as
\begin{subequations}
\begin{equation}
\label{energy_conservation}
    c_{heat}\sum_{k \in \mathbf{K}}{m_z^i\tilde{\tau}_{in,bc}^{t}=c_{heat}\sum_{u \in \mathbf{K}}{m_z^i\tau _{in,z}^{k,t}}},
\end{equation}
wherein $\mathbf{K}=[1, \cdots. K]$, $c_{heat}$ is the heat capacity of air, and $m_z^k$ is the air mass of agent $i$. 

Denote $\xi_{bc}^i=m_z^i/\sum_{i\in \mathbf{K}} m_z^i$ as the aggregation coefficient of the zone $i$. Then, we can define the aggregate state based on (\ref{energy_conservation}) using the aggregation equation \cite{lu2021data} as
\begin{equation}
    \label{aggregation_equation}
    \tilde{\tau}_{in,bc}^{t}=\sum_{i \in \mathbf{K}}{\xi _{bc}^{i}\tau _{in,z}^{i,t}}.
\end{equation}

According to the physical meaning of the aggregation coefficient given in (\ref{energy_conservation}), $\xi_{bc}^i$ satisfies 
\begin{equation}
    \label{aggregation coefficient equation}
    \sum_{i \in \mathbf{K}}{\xi _{bc}^{i}=1}, \ \ \xi_{bc}^i \geq 0, \ \forall i \in \mathbf{K}.
\end{equation}

Finally, by using the aggregate state, $\tilde{\tau}_{in,bc}^t$, to represent the state of the building cluster
, we get the ATDM of the building cluster as
\begin{equation}
    \label{ATDM}
    \begin{aligned}
        \tilde{\tau}_{in,bc}^{t}&=\sum_{m \in \mathbf{M} \backslash \{0\}}{\alpha _{bc}^{m}\tilde{\tau}_{in,bc}^{t-m} +
        \sum_{m \in \mathbf{M}}\sum_{i \in \mathbf{K}}{\beta _{bc}^{m}h_{load,z}^{i,t-m}}} \\ 
        &+\sum_{m \in \mathbf{M}}{\gamma _{bc}^{m}\tau _{out}^{t-m}}+\sum_{m \in \mathbf{M}}{\theta _{bc}^{m}h_{rad}^{t-m}} \\ 
        &+\tau _{occ,bc}^{t},\ \forall t \in \mathbf{T},
    \end{aligned}
\end{equation}
This model describes the relationship between the heating/cooling power $h_{load,z}^{i,t}$ and the aggregate state $\tilde{\tau}_{in,bc}^t$ of the building cluster, which can be directly used by energy systems for operation and control.
\end{subequations}

\vspace{-0.2cm}
\subsection{Parameter Estimation Model}
Combining the aggregation equation (\ref{aggregation_equation}) and the ATDM (\ref{ATDM}), we obtain the measurement equation as
\begin{equation}
    \label{measurement equation}
    \begin{aligned}
        \sum_{i \in \mathbf{K}}{\xi _{bc}^{i}\tau _{in,z}^{i,t}} &=
        \sum_{m \in \mathbf{M} \backslash \{0\}}{\sum_{i \in \mathbf{K}}{\alpha _{bc}^{m}\xi _{bc}^{i}\tau _{in,z}^{i,t-m}}} \\
        &+\sum_{m \in \mathbf{M}}\sum_{i \in \mathbf{K}}{\beta _{bc}^{m}h_{load,z}^{i,t-m}} +\sum_{m \in \mathbf{M}}{\gamma _{bc}^{m}\tau _{out}^{t-m}} \\
        &+\sum_{m \in \mathbf{M}}{\theta _{bc}^{m}h_{rad}^{t-m}} +\tau _{occ,bc}^{t} + \varepsilon^t, \ \forall t \in \mathbf{T}.
    \end{aligned}
\end{equation}

Here, given \{$\tau_{in,z}^{i,t-m}$, $h_{load,z}^{i,t-m}$, $\tau_{out}^{t-m}$, $h_{rad}^{t-m}$\} ($\forall m \in \mathbf{M},i \in \mathbf{K},t\in \mathbf{T}$), we can use the least squares regression (LSR) method to estimate the model parameters \{$\xi_{bc}^i$, $\alpha_{bc}^{m}$, $\beta_{bc}^{m}$, $\gamma_{bc}^{m}$, $\theta_{bc}^{m}$, $\tau_{occ,bc}^{t}$\} ($\forall m \in \mathbf{M}, i \in \mathbf{K}, t\in \mathbf{T}$) for independent and identically distributed Gaussian errors, $\varepsilon^t, t\in \mathbf{T}$. Then, let us denote $\xi=\left[\xi _{bc}^{1},\cdots, \xi _{bc}^{K}\right]^T \in \mathbb{R}^{K\times1}$, $\alpha=\left[\alpha_{bc}^1,\cdots,\alpha_{bc}^M\right]^T\in \mathbb{R}^{M\times 1}$, $\beta=\left[\beta_{bc}^0,\cdots,\beta_{bc}^M\right]^T\in \mathbb{R}^{(M+1)\times 1}$,  $\gamma=\left[\gamma_{bc}^0,\cdots,\gamma_{bc}^M\right]^T\in \mathbb{R}^{(M+1)\times 1}$, $\theta=\left[\theta_{bc}^0,\cdots,\theta_{bc}^M\right]^T\in \mathbb{R}^{(M+1)\times 1}$, $\tau_{occ}=\left[\tau_{occ}^1,\cdots,\tau_{occ}^{T}\right]^T$ $\tau_{out}^{-m} = \left[\tau _{out}^{1-m}, \cdots, \tau _{out}^{T-m}\right]^T\in \mathbb{R}^{T\times 1}$,  $h_{rad}^{-m}=\left[h_{rad}^{1-m}, \cdots, h_{rad}^{T-m}\right]^T\in \mathbb{R}^{T\times 1}$, 
\begin{align*}
    h_{load,z}^{-m}=\left[ \begin{matrix}
	h_{load,z}^{1,1-m}&		\cdots&		h_{load,z}^{K,1-m} \\
	\vdots&		\ddots&		\vdots\\
	h_{load,z}^{1,T-m}&		\cdots&		h_{load,z}^{K,T-m}
\end{matrix} \right]\in \mathbb{R}^{T\times K},
\end{align*}
\begin{align*}
    \tau _{in,z}^{-m}=\left[ \begin{matrix}
	\tau _{in,z}^{1,1-m}&		\cdots&		\tau _{in,z}^{K,1-m}\\
	\vdots&		\ddots&		\vdots\\
	\tau _{in,z}^{1,T-m}&		\cdots&		\tau _{in,z}^{K,T-m}\\
\end{matrix} \right]\in \mathbb{R}^{T\times K}. 
\end{align*}

Also,  to avoid the sparsity of the aggregation coefficients, $\xi$, and to address the inherent colinearity among the indoor temperature, we further add the $L_2$ regularization term in the LSR as
\begin{equation}
    \label{parameter_estimation_model_for_ATDM}
    \begin{gathered}
        \underset{\xi, \alpha, \beta, \gamma, \tau_{occ}}{\min}f(\xi, \alpha, \beta, \gamma, \theta,\tau_{occ}) \\
        =\left\| c_0\xi-c_1(\mathrm{I}_M \otimes \xi)\alpha
        - c_2 \beta
        - c_3 \gamma  \right.\\
        \left.- c_4 \theta 
        - \tau_{occ}\right\|_2^2 
        +\lambda\left\|\xi \right\|_2^2\\
       \textrm{s.t}.\ \  \xi  \geq 0, \  1^T\xi=1,
    \end{gathered}
\end{equation}
wherein $f(\cdot)$ is the objective function accounting for the sum of squared residuals as well as $L_2$ regularization term; $\lambda$ is the penalty factor; $\mathrm{1}_K$ is the $K$-dimensional $1$-vector;  $\mathrm{I}_x$ is the $x$-dimensional identity matrix; and $\otimes$ is the Kronecker product.
The constants coefficient matrices  $c_0$-$c_4$ are defined as
$c_0 = \tau_{in,z}^{-0}\in \mathbb{R}^{T\times K}$, $c_1 = [\tau_{in,z}^{-1},\cdots,\tau_{in,z}^{-M}]\in \mathbb{R}^{T \times KM}$, $c_2=[h_{load,z}^{-0}1_K,\cdots,h_{load,z}^{-M}1_K]\in \mathbb{R}^{T\times (M+1)}$, $c_3=[\tau^{-0}_{out},\cdots,\tau^{-M}_{out}]\in \mathbb{R}^{T\times(M+1)}$, 
$c_4=[h_{rad}^{-0},\cdots,h_{rad}^{-M}]\in \mathbb{R}^{T\times (M+1)}$; and $\otimes$ is the Kronecker product.


\begin{remark}
The model (\ref{parameter_estimation_model_for_ATDM}) is a nonconvex optimization problem with a nonconvex objective and linear constraints. Currently, there are no computationally effective methods to obtain the global optimum. Fortunately, one reasonable local optimum is acceptable for the parameter estimation problem. Besides, the existing research on privacy-preserved computation mainly focuses on convex optimization problems, such as linear and quadratic optimization, which cannot be directly applied to this problem.
\end{remark}

\section{Privacy-Preserved Algorithm}
In this section, we propose the privacy-preserved algorithm for the model (\ref{parameter_estimation_model_for_ATDM}). First, we provide the preliminaries on the privacy definition and related methods. Second, we use the BCD method to decouple the model (\ref{parameter_estimation_model_for_ATDM}) into two convex subproblems to tackle the non-convexity. Then, we present the privacy-preserved computation methods for the subproblems. 
\subsection{Preliminaries}
\subsubsection{Privacy Definition}
    The indoor temperature, $\tau_{in,z}^{i,t}$, ($\forall i \in \mathbf{K},\ \forall t \in \mathbf{T}$, and heating/cooling power, $h_{load,z}^{i,t}$, ($\forall i \in \mathbf{K},\ \forall t \in \mathbf{T}$), are defined as the private information of the user $i$.

\subsubsection{Secure Aggregation Protocol}
The SAP involves two types of entities: a computing center that performs aggregation operations and a set $\mathbf{U}$ composed of several users, where each user $i$ owns its private information, $x_i\in \mathbb{R}^{T\times 1}$. To guarantee that the computing center only acquires the summation, $\sum_{i\in \mathbf{U}} x_i$, without knowing private information, $x_i$, the following steps are implemented:

(1) Each pair of users $(i,j)$ shares a random vector $s_{i,j}\in \mathbb{R}^{T\times 1}$.

(2) User $i$ masks its private information, $x_i$, with the random vectors it receives to get the masked information, $\tilde{x}_i$, as

\begin{subequations}
\begin{equation}
    \label{SAP}
    \tilde{x}_i=x_i+\sum_{j\in \mathbf{U},j>i}{s_{i,j}}-\sum_{j\in \mathbf{U},j<i}{s_{j,i}}.
\end{equation}

(3) Each user sends the masked information, $\tilde{x_i}$, to the computing center, and the computing center calculates the summation of the masked information, denoted as $z$ as

\begin{equation}
    \label{SAP_sum}
    \begin{aligned}
        z&=\sum_{i\in \mathbf{U}}{\tilde{x}_i} 
        =\sum_{i\in \mathbf{U}}{x_i}.
    \end{aligned}
\end{equation}
\end{subequations}

Note that the SAP can be extended to the situation where $x_i$ is a matrix of any dimension as long as the random $s_{i,j}$ keeps the same dimension as $x_i$. 

\subsubsection{Transformation-based Encryption}
We take the unconstrained linear programming problem among multi-agents as an example to illustrate  the TE method, defined as

\begin{subequations}
\begin{equation}
\label{TE_ori_pro}
    \begin{gathered}
        \underset{x}{\min}\ c^Tx,
    \end{gathered}
\end{equation}
wherein $x=[x_1,\cdots,x_K]^T\in \mathbb{R}^{T\times 1}$, $c=[c_1, \cdots
,c_{K}]^T\in \mathbb{R}^{K \times 1}$.
In this problem, $c_i$ is the private information belonging to agent $i$,  and $x_i$ is its associated decision variable. The computing center needs $c$ to implement the optimization calculation. The fundamental idea of the TE method is to mask the input data through a random matrix, including additive randomization and multiplicative randomization. In this work, multiplicative randomization is used, and the following steps are implemented.  

(1) Each agent generates a random column vector $W^{[i]}=\left[w_{1i},\cdots,w_{Ki}\right]^T\in \mathbb{R}^{K\times 1}$ as the encryption matrix, only known to itself. Denote $W=\left[W^{[1]},\cdots,W^{[K]}\right]\in \mathbb{R}^{K\times K}$.

(2)  Each agent uploads the product, $c_i (W^{[i]})^T$, to the computing center. Then, the computing center calculates the summation, $\sum_{i\in \mathbf{K}}{c_i(W^{[i]})^T}$, and constructs the equivalent optimization problem in \eqref{TE_equi_pro},
\begin{equation}
\label{TE_equi_pro}
    \begin{gathered}
        \underset{\bar{x}}{\min}\ c^TW^T\bar{x},
    \end{gathered}
\end{equation}
where $\bar{x}\in \mathbb{R}^{K\times 1}$ is the encrypted decision vector.

(3) The computing center solves the optimization problem \eqref{TE_equi_pro} to get $\bar{x}$, and then broadcasts $\bar{x}$ to all agents. Then, the agent $i$ can recover its decision variable according to the recovery equation as
\begin{equation}
    \label{recovery_equation}
    x_i=(W^{[i]})^T\bar x.
\end{equation}
\end{subequations}

\begin{remark}
The TE technique can be applied to the parameter estimation problem by taking the decision variables in \eqref{TE_ori_pro} as the parameters to be estimated. 
\end{remark}

\subsection{The BCD-Based Iterative Algorithm}
The existing privacy-preserved computation methods usually cannot be directly applied to the model (\ref{parameter_estimation_model_for_ATDM}) because of the nonconvex objective. Fortunately, this problem is easy to decompose. Specifically, the model (\ref{parameter_estimation_model_for_ATDM}) turns into a quadratic programming problem once the variables $\xi$ or $\alpha$ are fixed, for which some existing privacy-preserved computation techniques can be used. Inspired by this, we use the BCD method to decouple the model (\ref{parameter_estimation_model_for_ATDM}) into a $\xi$-fixed problem and a $\alpha$-fixed problem as 
\begin{subequations}
\begin{equation}
    \label{sub_problem_I}
    \begin{gathered}
        {\text{SP}_\text{I}}(\xi): \ \
        \underset{\alpha, \beta, \gamma, \theta, \tau_{occ}}{\min}f(\xi, \alpha, \beta, \gamma, \theta, \tau_{occ}),
    \end{gathered}
\end{equation}

\begin{equation}
    \label{sub_problem_II}
    \begin{gathered}
        {\text{SP}_\text{II}}(\alpha): \ \
        \underset{\xi, \beta, \gamma, \theta, \tau_{occ}}{\min} f(\xi, \alpha, \beta, \gamma, \theta, \tau_{occ})\\
        \quad \quad \quad \quad \textrm{s.t}.\ \ \xi \geq 0,\ 1^T\xi=1.
    \end{gathered}
\end{equation}
\end{subequations}

Theoretically, the local optimum of the model (\ref{parameter_estimation_model_for_ATDM}) can be obtained by solving (\ref{sub_problem_I}) and (\ref{sub_problem_II}) iteratively. As indicated in (\ref{parameter_estimation_model_for_ATDM}), the privacy information in the two subproblems includes $c_0$, $c_1$, and $c_2$. 
We only focus on protecting $c_0$ and $c_1$ in the following for conciseness, since $c_2$ can be protected the same way as section \ref{Privacy-Preserved Computation for sp1} does through the SAP method.

\subsection{Privacy-Preserved Computation for $SP_I(\xi)$} 
\label{Privacy-Preserved Computation for sp1}
In $\mathrm{SP}_\mathrm{I}(\xi)$, since $\xi$ is known and can be transferred to the users, the privacy of $c_0$ and $c_1$ equals that of $c_0 \xi$ and $c_1(\mathrm{I}_M \otimes \xi)$. Note that $c_0 \xi$ and $c_1(\mathrm{I}_M \otimes \xi)$ have a summation form, so we propose the SAP-based computation procedure as follows. 

\begin{subequations}
\begin {enumerate}
\item{The BLA transfers $\xi_i$ to the agent $i$, $\forall i \in \mathbf{K}$.}
\item{The agent $i$ calculate $s_i^{-m}$, as}
\begin{equation}
    \label{s_i_definition}
    s_i^{-m}=\xi _i (\tau_{in,z}^{-m})^{[i]}, \ \forall i\in \mathbf{K}, \forall m\in \mathbf{M},
\end{equation}
\item{The agent $i$ generates random vectors $r_{i,j}^{-m}\in \mathbb{R}^{T\times 1}, \forall m \in \mathbf{M}, \forall i \in \mathbf{K}$, and transfer them to $j, \forall i < j \leq K$. Afterwards, the agent $i$ perform the following computation}
\begin{equation}
    \label{SAP_s_i}
    \tilde{s}_i^{-m}=s_i^{-m}+\sum_{j>i}{r_{i,j}^{-m}}-\sum_{j<i}{r_{j,i}^{-m}}.
\end{equation}
\item{The agent $i$ send the masked private information $\tilde{s}_i^{-m}$ to the BLA, $\forall i \in \mathbf{K}$.}
\item{The BLA calculates $c_0\xi$ and $c_1(\mathrm{I}_M \otimes \xi)$ using $\tilde{s}_i^{-m}, \forall i\in \mathbf{K}, m \in \mathbf{M}$, as}
\begin{equation}
    \label{SAP_s_i_sum}
    \begin{aligned}
    c_0\xi &= \tau_{in,z}^{-0}\xi = \sum_{i \in \mathbf{K}}{s_i^{-0}}=\sum_{i \in \mathbf{K}}{\tilde{s}_i^{-0}},
    \end{aligned}   
\end{equation}
\begin{equation}
    \label{SAP_s_i_sum_2}
    \begin{aligned}
    c_1(\mathrm{I}_M \otimes \xi) &= \left[\tau_{in,z}^{-1}\xi,\cdots,\tau_{in,z}^{-M}\xi \right] \\
    &=\left[\sum_{i \in \mathbf{K}}{\tilde{s}_i^{-1}},\cdots,\sum_{i \in \mathbf{K}}{\tilde{s}_i^{-M}}\right],
    \end{aligned}   
\end{equation}
and then solve $\mathrm{SP}_\mathrm{I}(\xi)$ to obtain $\alpha(\xi)$, $\beta(\xi)$, $\gamma(\xi)$, $\theta(\xi)$, and $\tau_{occ}(\xi)$.
\end{enumerate}
\end{subequations}

\begin{remark}
Obviously, the SAP method also applies to $c_2$ since its element has the summation form, i.e., $h_{load,z}^{-m}{1_K}$, in which the agent $i$ only needs to generate extra random vectors to mask $(h_{load,z}^{-m})^{[i]}, \forall m\in \mathbf{M}$. Hence, we do not detail the privacy preservation method of $c_2$ for conciseness.
\end{remark}





\subsection{Privacy-Preserved Computation for $SP_{II}(\alpha)$} 
We assume that the BLA transfers $\alpha$ to each agent. Define
\begin{subequations}
\begin{equation}
\label{hat_tau_definition}
\begin{aligned}
\hat{\tau}_{in,z} \triangleq \tau_{in,z}^{-0}-\sum_{m \in \mathbf{M \backslash \{0\}}}\alpha_{bc}^{m} \tau_{in,z}^{-m} \in \mathbb{R}^{T\times K},
\end{aligned}
\end{equation}
and then we have
\begin{equation}
    \label{new form of f(x)}
    \begin{gathered}
        f(\xi, \alpha, \beta, \gamma, \theta, \tau_{occ})
        =\left\| 
        \hat{\tau}_{in,z}\xi - c_2 \beta
        - c_3 \gamma\right. \\
        \left.- c_4 \theta 
        - \tau_{occ} 
        \right\|_2^2 
        +\lambda\left\|\xi \right\|_2^2.
    \end{gathered}
\end{equation}
\end{subequations}

After the agent $i$ receives $\alpha$, it can calculate $\hat{\tau}_{in,z}^{[i]}$ that is needed to solve $\mathrm{SP}_\mathrm{II}(\alpha)$. However, if the BLA knows $\hat{\tau}_{in,z}$, it can infer the privacy information $\tau_{in,z}^{-m} \ (\forall m\in \mathbf{M}$) of users. Our detailed proof of this is given in Appendix \ref{appendix_a}. Hence, in the following, we propose the TE-based method to mask $\hat{\tau}_{in,z}$.

We assume that $\xi$ can be obtained based on $\bar{\xi}$ through a linear transformation as
\begin{subequations}
\begin{equation}
    \label{transformation_based_encryption_xi}
    \xi =W^T\bar{\xi},
\end{equation}
wherein
\begin{equation}
    \label{W_definition}
    W=\left[\begin{matrix}
        W^{[1]}& \cdots& W^{[K]}
    \end{matrix}
    \right]=
    \left[ \begin{matrix}
	w_{11}&		\cdots&		w_{1K}\\
	\vdots&		\ddots&		\vdots\\
	w_{K1}&		\cdots&		w_{KK}\\
\end{matrix} \right]. 
\end{equation}
\end{subequations}

In \eqref{W_definition}, $W^{[i]}$ is the random matrix belonging to the agent $i$. 
Then, $\text{SP}_\text{II}(\alpha)$ can be represented as
\begin{equation}
    \label{masked_SP_II}
    \begin{gathered}
     \underset{\bar{\xi}, \beta, \gamma, \theta, \tau_{occ}}{\min}
     f(\bar{\xi}, \beta, \gamma, \theta, \tau_{occ}) =\left\| 
        \hat{\tau}_{in,z}W^T \bar{\xi} - c_2 \beta \right. \\
        \left. - c_3 \gamma - c_4 \theta 
        - \tau_{occ} 
        \right\|_2^2 
        +\lambda\bar\xi^TWW^T\bar\xi\\
       \textrm{s.t}.\ \ W^T \bar{\xi} \geq 0, 1^TW^T \bar{\xi}=1.
    \end{gathered}
\end{equation}

To solve (\ref{masked_SP_II}), the BLA needs to know the three uploads from agents, namely $\hat{\tau}_{in,z} W^T$, $W W^T$, and $1^T W^T$. Note that $W$ cannot be known by the BLA; otherwise, the BLA can probably infer $\hat{\tau}_{in,z}$ by combining $\hat{\tau}_{in,z} W^T$ and $W^T$, and can further $\tau_{in,z}^{i,t}\ (\forall i\in \mathbf{K}, \forall t\in \mathbf{T})$. Therefore, each agent $i$ cannot directly upload $W^{[i]}$ to the BLA. This is where the SAP method comes into play. We notice $\hat{\tau}_{in,z} W^T$, $W W^T$, and $1^T W^T$ can be reformulated into summation forms as
\begin{subequations}
    \begin{equation}
    \label{hat_tau_partitioned}                       
    \hat{\tau}_{in,z} W^T =\sum_{i\in \mathbf{K}}\hat{\tau}_{in,z}^{[i]}(W^{[i]})^T,
    \end{equation}
    \begin{equation}
    \label{W.W^T}                       
     WW^T = \sum_{i\in \mathbf{K}}W^{[i]} (W^{[i]})^T.
    \end{equation}
    \begin{equation}
    \label{1^T.W^T}                       
     1^TW^T = \sum_{i\in \mathbf{K}}(W^{[i]})^T,
    \end{equation}

\end{subequations}

Hence, by denoting $A_1^i=\hat{\tau}_{in,z}^{[i]}(W^{[i]})^T \in \mathbb{R}^{T \times K}$ and $A_2^i= W^{[i]}(W^{[i]})^T \in \mathbb{R}^{K \times K}$, the SAP method can be implemented as
\begin{subequations}
    \begin{equation}
    \label{SAP_A1}
    \tilde{A}_1^i=A_1^i +\sum_{j\in \mathbf{K},j>i}{u_{i,j}-\sum_{j\in \mathbf{K},j<i}{u_{j,i}}},
\end{equation}
\begin{equation}
    \label{SAP_A2}
    \tilde{A}_2^{i}=A_2^{i} + \sum_{j\in \mathbf{K},j>i}p_{i,j}-\sum_{j\in \mathbf{K},j<i}p_{j,i},
\end{equation}
\begin{equation}
    \label{SAP_W}
    \tilde{W}^{[i]}= W^{[i]} + \sum_{j\in \mathbf{K},j>i}q_{i,j}-\sum_{j\in \mathbf{K},j<i}q_{j,i},
\end{equation}
\end{subequations}
where $u_{i,j} \in \mathbb{R}^{T\times K}$, $p_{i,j} \in \mathbb{R}^{K\times K}$, and $q_{i,j}\in \mathbb{R}^{K\times 1}$ are the random matrices generated by the agent $i$.

 It is worth pointing out that the agent $i$ cannot directly upload $A_1^i$ and $A_2^i$ to the BLA. Once it occurs, the BLA can infer the proportional relationship of each element in $M^{[i]}$, and further infer $w_{i,j}\ (\forall i,j\in \mathbf{K})$ based on $\xi=W^T\bar\xi$. To illustrate this, we take what will happen when $A_2^i$ is disclosed to the BLA as an example. The details can be found in Appendix \ref{appendix_b}.
Besides, the constraint $W^T \bar{\xi} \geq 0$ in (\ref{masked_SP_II}), originated from $\xi \geq 0$, is a tricky problem since the BLA cannot know the value of $W$. Fortunately, the simulation results indicate $\xi \geq 0$ always holds even if we remove it from $\mathrm{SP}_\mathrm{II}(\alpha)$. Physically, this constraint ensures that the contribution of each zone to the aggregate state is not negative, which is obviously established.

Finally, we obtain the transformed model for $\mathrm{SP}_\mathrm{II}(\alpha)$ as
\begin{equation}
    \label{masked_SP_II_final}
    \begin{gathered}
     \underset{\bar{\xi}, \beta, \gamma, \theta, \tau_{occ}}{\min}
     f(W^T\bar{\xi}, \alpha, \beta, \gamma, \theta, \tau_{occ}) =\left\| 
        \textstyle \left(\sum_{i \in \mathbf{K}} \tilde{A_1^i}\right) \bar{\xi} - c_2 \beta \right. \\
        \left. - c_3 \gamma - c_4 \theta 
        - \tau_{occ} 
        \right\|_2^2 
        +\lambda\bar\xi^T \textstyle \left(\sum_{i \in \mathbf{K}} \tilde{A_2^i}\right) \bar\xi\\
       \textrm{s.t.} \ \ \textstyle \left(\sum_{i \in \mathbf{K}} \tilde{W}^{[i]}\right) \bar{\xi}=1.
    \end{gathered}
\end{equation}

Now, the privacy-preserved computation procedure for $\mathrm{SP}_\mathrm{II}(\alpha)$ is given as follows.
\begin {enumerate}
\item{The BLA broadcast $\alpha$ to each agent.}

\item{The agent $i$ calculates $\hat{\tau}_{in,z}^{[i]}$ based on (\ref{hat_tau_definition}), $\forall i\in \mathbf{K}$.}

\item{The agent $i$ generates random matrices (vectors) $W^{[i]}$, $u_{i,j}$, $p_{i,j}$, and $q_{i,j}$, and transfers $u_{i,j}$, $p_{i,j}$, and $q_{i,j}$ to other zones. ($i<j\leq K$)}.

\item{The agent $i$ calculates $\tilde{A}_1^{[i]}$, $\tilde{A}_2^{[i]}$, and $\tilde{W}^{[i]}$ based on (\ref{SAP_A1})-(\ref{SAP_W}), and then transfers them to the BLA.}

\item{The BLA solves the problem (\ref{masked_SP_II_final}) to obtain $\bar{\xi}(\alpha)$, $\beta(\alpha)$, $\gamma(\alpha)$, $\theta(\alpha)$, and $\tau_{occ}(\alpha)$, and then broadcasts $\bar{\xi}(\alpha)$ to each agent.}

\item{The agent $i$ calculates $\xi_{bc}^i = \left(W^{[i]}\right)^T \bar{\xi}(\alpha)$, and then return $\xi_{bc}^i$ to the BLA.}

\end{enumerate}

\subsection{Flowchart of the Proposed Algorithm}
\begin{algorithm}[t]
    \SetAlgoLined 
	\caption{Privacy-preserved Algorithm for ATDM.}
	\KwIn{$h_{load,z}^{-m}$, $\tau_{out}^{-m}$, $h_{rad}^{-m},\ \forall m\in \mathbf{M}$}
	\KwOut{the final solution and the objective function}
	The BLA set tolerance $\delta$, initialize $\xi^{(0)}$, $l=0$, and $GAP=+\infty$;  the BLA distributes $\xi_i^{(0)}$ to agent $i$\; 
	\While{$GAP\geq \delta$}{
        Each agent $i$ generates random vector $r_{i,j}^{-m,(l)}$, sends it to agent $j$, and calculates $\tilde s_i^{-m,(l)}$ and sends it to the BLA\;
        The BLA calculates $\sum_{i=1}^K \tilde s_i^{-m,(l)}$ , solve $\mathrm{SP}_{\mathrm{I}}(\xi)$, get the optimal solution $f_1$ and $\alpha^{(l+1)}$\;
        The BLA sends $\alpha^{(l+1)}$ to each agent \;
        Each agent $i$ generates random matrices $W^{[i],(l)}$, $u_{i,j}^{(l)}$, $p_{i,j}^{(l)}$ and $q_{i,j}^{(l)}$. Each agent $i$ calculates $\tilde{A}_1^{i,(l)}$, $\tilde{A}_2^{i,(l)}$ and $\tilde{W}^{[i],(l)}$ separately and sends them to the BLA\;
        The BLA solve the transformed $\mathrm{SP}_{\mathrm{II}}(\alpha)$ in \eqref{masked_SP_II_final}, get the optimal solution $f_2$ and $\bar{\xi}^{(l+1)}$\; 
        The BLA distributes $\bar{\xi}^{(l+1)}$ to the agent $i$\; 
        Each agent $i$ calculates $\xi_{bc}^{i,(l+1)}$ according to $\xi_{bc}^{i,(l+1)}=(W^{[i],(l)})^T\bar{\xi}^{(l+1)}$, and sent $\xi_{bc}^{i,(l+1)}$ back to the BLA\;
        The BLA calculates the $GAP$: 
        \begin{align*}
            GAP=\min \left\{f_1-f_2,(f_1-f_2)/f_2\right\};\;
        \end{align*}
        $l=l+1$\;
	}
        \label{Algoritm 1}
\end{algorithm}

\begin{figure}[t]
 \centering
 \footnotesize
 \includegraphics[width=1\linewidth]{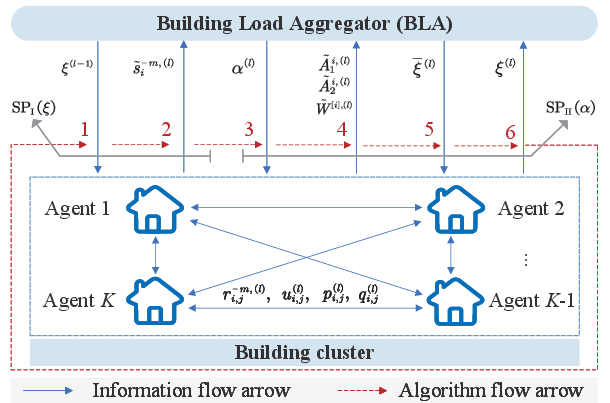}
 \caption{Information exchange of the proposed algorithm. The blue arrows between the BLA block and the building cluster block represent the interactive information within these two entities. The blue arrows within the building cluster block denote the intermediate information between two agents that is necessary for SAP. The red arrows represent the algorithm procedures performed between information transformations, among which the first two belong to $\mathrm{SP}_\mathrm{I}(\xi)$ and the others belong to $\mathrm{SP}_\mathrm{II}(\alpha)$.}
 \label{fig_Information_exchange}
 \end{figure}

The flowchart of the proposed privacy-preserved algorithm is given in Algorithm 1. The information exchange process of the algorithm is shown in Fig. \ref{fig_Information_exchange}. Here, we suppose that the agents are independent entities that can communicate with other agents and the BLA. The algorithm can be easily ported to the system between the buildings and the BLA by considering the buildings as independent entities. Note that in the proposed algorithm, the BLA needs to solve the variants of $\mathrm{SP}_\mathrm{I}(\xi)$ and $\mathrm{SP}_\mathrm{II}(\alpha)$ in Step 4 and Step 7, respectively, which are both quadratic programming problem that can be solved by many off-the-shelf software.

\section{Privacy Analysis}
In this section, privacy analysis will be carried out for the proposed privacy-preserved ATDM. First, 
for the BLA to make privacy inferences, it has to aggregate the received information to offset the random matrices introduced by the SAP method. Second, we divide the effective information for privacy inferences into three categories, i.e., only $\tau$-related information, only $W$-related information, and $\{\tau, W\}$-related information. Then, we analyze the impossibility of inferring privacy information based on only $\tau$-related or only $W$-related information. Finally, we clarify that the essence of privacy inferences in our proposed algorithm is to solve the MQS problem, which is NP-hard. We start with a lemma to illustrate the privacy security of the SAP method.

\begin{lemma}
\label{lemma_1}
In the SAP method, random masks hide all information about users’ individual privacy information except for their sum.
\end{lemma}

We omit the proof of Lemma 1. The relative analysis can be found in \cite{bonawitz2017practical}.

\begin{remark}
Lemma 1 conveys a basic idea that once the BLA tends to make privacy inferences, it has to aggregate the information uploaded by agents to eliminate the substantial random variables introduced by SAP. Thus, the BLA needs to aggregate the received information $\tilde{s}_i^{-m} (\forall i\in \mathbf{K}, \forall m\in \mathbf{M})$ in $\mathrm{SP}_{\mathrm{I}}(\xi)$ and 
$\tilde{A}_1^i$, $\tilde{A}_2^i$ as well as $\tilde{W}^{[i]}$ ($\forall i\in \mathbf{K}$) in $\mathrm{SP}_{\mathrm{II}}(\alpha)$ to get aggregate information $\tau_{in,z}^{-m}\xi \ (\forall m\in \mathbf{M})$, $\hat{\tau}_{in,z}W^T$, $WW^T$ and $1^TW^T$ for privacy inferences. Besides, the relationship between $\xi$ and $\bar{\xi}$, i.e., $\xi=W^T\bar{\xi}$, is also valuable for the BLA to make privacy inferences. This relationship information is caused by implementing the 7th step of our proposed algorithm.
\end{remark}


 Denote  $\tau_{in,z}^{-m}\xi^{(l)}$ as $d_{1,m}^{(l)}\in \mathbb{R}^{T\times 1}$, $(W^{(l)} W^{(l)})^T$ as $D_1^{(l)}\in \mathbb{R}^{K\times K}$,  $1^T (W^{(l)})^T$  as $d_2^{(l)}\in \mathbb{R}^{K\times 1}$ and $\hat{\tau}_{in,z}(W^{(l)})^T$ as $D_2^{(l)}\in \mathbb{R}^{T\times K}$. Then, We divide the effective information for privacy inferences into three types:
\begin{enumerate}
    \begin{subequations}
    \item Type-1 information: only $\tau$-related information 
    \begin{equation}
        \mathrm{I}^{(l)}_{\tau}\triangleq \left \{\tau_{in,z}^{-m} \ | \ \tau_{in,z}^{-m}\xi^{(l)}=d_{1,m}^{(l)} , m\in \mathbf{M} \right \}
    \end{equation}
    \item Type-2 information: only $W$-related information 
    \begin{equation}
        \begin{gathered}
        \mathrm{I}^{(l)}_W\triangleq \left\{W^{(l)} \ |\ \right. W^{(l)}(W^{(l)})^T=D_1^{(l)},\\
        1^T(W^{(l)})^T=d_2^{(l)},\left.(W^{(l)})^T\bar{\xi}^{(l)} = \xi^{(l)} \right\}
        \end{gathered}
    \end{equation}
    \item Type-3 information: $\{\tau, W\}$-related information 
    \begin{equation}
    \begin{gathered}
         \mathrm{I}^{(l)}_{\tau,W}\triangleq \left\{\tau_{in,z}^{-m}, W^{(l)}, m\in \mathbf{M} \ | \ \right.\\ 
         \left. (\tau_{in,z}^{-0}-\sum_{m \in \mathbf{M \backslash \{0\}}}\alpha_{bc}^{m} \tau_{in,z}^{-m})(W^{(l)})^T = D_2^{(l)}\right\}
    \end{gathered}
    \end{equation}
    \end{subequations}
    
\end{enumerate}
wherein $l\in \mathbf{L}$ is the iteration index. Type-1 information is only useful for inferring indoor temperatures, $\tau_{in,z}^{-m}$; Type-2 information is only useful for inferring the random matrix,  $W^{(l)}$; Type-3 information plays roles in inferring both $\tau_{in,z}^{-m}$ and $W^{(l)}$.

\begin{proposition}
\label{proposition_1}
\textit{In the privacy preserved algorithm for ATDM proposed above, the BLA cannot infer the indoor temperature information $\tau_{in,z}^{i,t}\ (\forall i\in \mathbf{K}, \forall t\in \mathbf{T})$ based on type-1 information when $K>L$.}    
\end{proposition}

\begin{proof}
Based on the type-1 information, the inference equations for the BLA are
\begin{equation}
    \label{pro_1_1}
    \tau_{in,z}^{-m}\xi^{(l)}=d_{1,m}^{(l)}, \ \forall m \in \mathbf{M}.
\end{equation}
Here, \eqref{pro_1_1} is essentially the aggregate temperature information from period $1-M$ to $T$. Therefore, $\tau_{in,z}^{i,1-M},\cdots,\tau_{in,z}^{i,T} \ (\forall i\in \mathbf{K})$ are $(T+M)K$ unknown variables. Correspondingly, the total number of equations for inference is $(T+M)L$.  According to $K>L$, $(T+M)K>(T+M)L$ can be derived, making \eqref{pro_1_1} an under-determined equation system.
\end{proof}

Then, we consider type-2 information because the matrix $W$ also needs to be protected according to the previous discussion. Since all the parameters and variables in type-2 information are renewed during each iteration, we only need to consider whether $W$ can be inferred during one particular iteration. Thus, we dismiss the iteration index when analyzing type-2 information $\mathrm{I}_W^{(l)}$ for simplicity. Proposition \ref{proposition_2} is a detailed elaboration. 

\begin{proposition}
\label{proposition_2}
\textit{In the privacy-preserved algorithm in Section III, the BLA cannot infer the privacy inferences based on type-2 information when $K\geq 6$.}    
\end{proposition}

\begin{proof}
Based on the definition of type-2 information, the BLA can get the inference equations as
\begin{subequations}
    \begin{equation}
        \label{pro_2_1}
        WW^T=D_1,
    \end{equation}
    \begin{equation}
        \label{pro_2_2}
        1^TW^T=d_2,
    \end{equation}
    \begin{equation}
        \label{pro_2_3}
        \xi = W^T\bar \xi,
    \end{equation}
\end{subequations}
wherein the random matrix, $W$, has $K^2$ unknown variables. Note that the matrix, $D_1$, is  symmetric, so there is $\sum_{i=1}^K i$, i.e. $K(K+1)/2$ independent equations in \eqref{pro_2_1}. \eqref{pro_2_2} and \eqref{pro_2_3} provides $K$ equations, respectively. Considering \eqref{pro_2_1}-\eqref{pro_2_3} comprehensively, there are $\frac{1}{2}K^2+\frac{5}{2}K$ equations for inference. When the condition $K\geq 6$ holds, we have $\frac{1}{2}K^2+\frac{5}{2}K < K^2$, which means the equation system is under-determined and $W$ cannot be inferred \cite{jia2022chance}.
\end{proof}

Proposition \ref{proposition_1} and Proposition \ref{proposition_2} imply that the BLA has to resort to type-3 information to make privacy inferences. The privacy inference problem is to solve an MQS problem once the BLA resorts to type-3 information as

\begin{subequations}
    \begin{equation}
    \label{type-1}
        \tau_{in,z}^{-m}\xi^{(l)}=d_{1,m}^{(l)},\ \forall m\in \mathbf{M}, l \in \mathbf{L},\\
    \end{equation}
    \begin{equation}
        \label{type-21}
        W^{(l)}(W^{(l)})^T=D_1^{(l)}, \forall l \in \mathbf{L},\\
    \end{equation}
    \begin{equation}
        \label{type-22}
        1^T(W^{(l)})^T=d_2^{(l)},\forall l \in \mathbf{L},\\
    \end{equation}
    \begin{equation}
        \label{type-23}
        (W^{(l)})^T\bar{\xi}^{(l)} = \xi^{(l)}, \forall l \in \mathbf{L},\\
    \end{equation}
    \begin{equation}
        \label{type-3}
        (\tau_{in,z}^{-0}-\sum_{m \in \mathbf{M \backslash \{0\}}}\alpha_{bc}^{m} \tau_{in,z}^{-m})(W^{(l)})^T=D_2^{(l)}, \forall l \in \mathbf{L}.
    \end{equation}
    \label{MQS}
\end{subequations}

The equations \eqref{type-1}, \eqref{type-22}, and \eqref{type-23}  are linear equations, while the other two are quadratic ones, making \eqref{MQS} an MQS system.  It is well known that the MQS problem is NP-hard, for which no polynomial-time algorithms exist \cite{thomae2010solving, Tanaka2014EvaluatingST}.

In addition, \eqref{type-1}-\eqref{type-3} provide $(T+M)L$, $\frac{1}{2}(k^2+k)L$, $KL$, $KL$, and $TKL$ equations, respectively, while $\tau_{in,z}^{-m} \ (\forall m\in \mathbf{M})$ and $W^{[l]} \ (\forall l \in \mathbf{L})$ have $(T+M)K$ and $K^2L$ unknown variables, respectively. When the number of periods $T$ is larger than the total number of building zones, $K$, the number of equations in \eqref{MQS} is larger than the number of unknown variables, making \eqref{MQS} an over-determined system. In this situation, we can only obtain the least squares solution of this over-determined system, which is a typical nonconvex optimization problem. Note that there are usually numerous local optima for this nonconvex problem, making it impossible to determine whether the obtained solution is right. In summary, the BLA can't infer the private information of the users. We will conduct further analysis and verification in simulations.

\section{Simulation Results}
The data from the REFIT Smart Home dataset \cite{Firth2017} is used to verify the effectiveness of the proposed privacy-preserved nonlinear parameter estimation method. Seven buildings with relatively complete data, namely No. 1, 8, 10, 11, 13, 16, and 20, are selected for simulation. In this dataset, the time resolution is 30 minutes, and the number of records is 1440. We divide the training set and the testing set according to the proportion of 3 to 1, i.e. 1080 records for the training set and 360 records for the testing set. The period of occupants' activities is assumed to be 24h, and thus $T_{occ}$ is set to 48. Each element in the random matrix, $W$, is set to obey the normal distribution with a standard deviation of 0.1 and a mean of 0.1. We set the model order, $M$, as 2, and the penalty factor, $\lambda$, as 100 \cite{lu2021data}.

All the simulations are performed on a PC with an Intel i7 core and 32GB RAM. The programming is based on MATLAB R2022b and Yalmip \cite{Lofberg2004}. Gurobi 10.0 is used to solve the quadratic programming problem. The iteration tolerance in the proposed algorithm is set as $10^{-6}$.

In the following, we analyze the performance of the proposed privacy-preserved algorithm for ATDM from four aspects: calculation accuracy, masking performance, computation performance, and privacy security.

\subsection{Accuracy Analysis}
\begin{figure}[t]  
 \captionsetup{singlelinecheck=on}
  \footnotesize 
  \centering         
  \subfloat[]  
  {
      \label{fig_result_2_a}\includegraphics[width=1\linewidth]{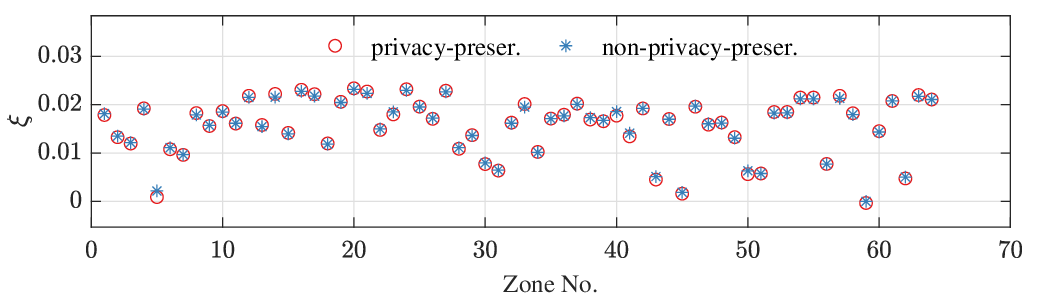}
  }\\
    \vspace{-0.3cm}
  \subfloat[] 
  {
      \label{fig_result_2_b}\includegraphics[width=1\linewidth]{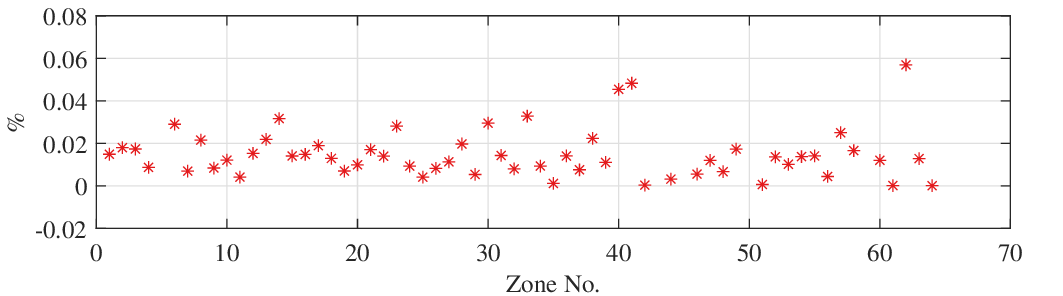}
  }   
 
  \caption{Aggregation coefficients: (a) Values (privacy-preserved vs. non-privacy-preserved algorithms); (b) Relative errors between privacy-preserved and non-privacy-preserved algorithms.}
  \label{fig_result_A_1}  
\end{figure}

The parameter estimation results by the privacy-preserved and non-privacy-preserved algorithms are given in Fig. \ref{fig_result_A_1} and Table \ref{table_result_A_1}. Obviously, the estimation results of the proposed privacy-preserved algorithm are highly close to the ones under the non-privacy-preserved algorithm, with the relative error smaller than 0.1\%. The minor errors are possibly caused by the rounding error of the random matrix introduced by the TE method. Thus, the calculation accuracy of the privacy-preserved algorithm for ATDM is well verified. 

\begin{table}[t]  
    \centering
    \footnotesize 
    \caption{Estimation Results of $\alpha$, $\beta$, $\gamma$, $\theta$}
    \vspace{-0.2cm}
    \label{table_result_A_1}
    \begin{tabular*}{\linewidth}{c c c c c}
    \toprule
    \makebox[0.15\linewidth][c]{\textbf{Method}} & \makebox[0.15\linewidth][c]{\textbf{Parameter}} & \makebox[0.15\linewidth][c]{\textbf{m=0}} & \makebox[0.15\linewidth][c]{\textbf{m=1}} &
    \makebox[0.15\linewidth][c]{\textbf{m=2}}  \\
    \midrule
    \multirow{4}{4em}{\textbf{Privacy-preserved}} & $\alpha_m$ & {/} & 1.4796 & -0.4884 \\ 
    & $\beta_m$ & 0.0295 & 0.0249 & 0.0042 \\
    & $\gamma_m$ & 0.0069 & -0.0057 & 0.0042 \\ 
    & $\theta_m$ & 0.3906 & 0.0692 & -0.1969 \\
    \midrule
    \multirow{4}{4em}{\textbf{Non-privacy-preserved.}} & $\alpha_m$ & {/} & 1.4804 & -0.4892 \\ 
    & $\beta_m$ & 0.0296 & 0.0247 & 0.0042 \\
    & $\gamma_m$ & 0.0066 & -0.0055 & 0.0040 \\ 
    & $\theta_m$ & 0.3904 & 0.0692 & -0.1972 \\
\bottomrule
\end{tabular*}
\end{table}

Then, we calculate some statistical indicators on the test set and compare them with the results without privacy protection to verify the accuracy of the algorithm. Here, three common statistical indicators are adopted: RMSE, MAPE, and $\text{R}^2$. Their calculation formulas are shown as 
\begin{subequations}
    \begin{equation*}
    \label{RMSE}    \mathrm{RMSE}=\sqrt{\frac{1}{T}\sum_{t=1}^{T}\left(\tilde{\tau}_{in,bc,pre}^{t}-\tilde{\tau}_{in,bc,real}^{t}\right)^2},
\end{equation*}
\begin{equation*}
    \label{MAPE}
    \mathrm{MAPE}=\frac{1}{T}\sum_{t=1}^{T}
    \mid \frac{(\tilde{\tau}_{in,bc,pre}^{t}-\tilde{\tau}_{in,bc,real}^{t})}{\tilde{\tau}_{in,bc,real}^{t}}  \mid, %
\end{equation*}
\begin{equation*}
    \label{R^2}
    \mathrm{R}^2=1-\frac{\sum_{t=1}^T \left( \tilde{\tau}_{in,bc,pre}^t-\tilde{\tau}_{in,bc,real}^t \right) ^2}{\sum_{t=1}^T\left(\tilde{\tau}_{in,bc,real}^t-1/T\sum_{t=1}^T\tilde{\tau}_{in,bc,real}^t\right)^2},
\end{equation*}
\end{subequations}
wherein $\tilde{\tau}_{in, bc, pre}^t$ is the predicted aggregate state of the building cluster calculated according to \eqref{ATDM} and $\tilde{\tau}_{in, bc, real}^t$ is the real aggregate state of the building cluster calculated according to  \eqref{aggregation_equation}.

The comparisons of these error indicators between the privacy-preserved and non-privacy-preserved algorithms are illustrated in Table \ref{table_result_A_2}. It can be found that the privacy-preserved algorithm can fit well with RMSE, MAPE, and $\text{R}^2$ equal to 0.2944℃, 1.3103\%, and 0.8613, respectively. Compared with the non-privacy-preserved algorithm, the accuracy of forecasting slightly decreased. In other words, the proposed privacy-preserved algorithm sacrifices a little prediction accuracy for excellent privacy-preservation performance.

 \begin{figure}[t]
 \centering
 \includegraphics[width=1\linewidth]{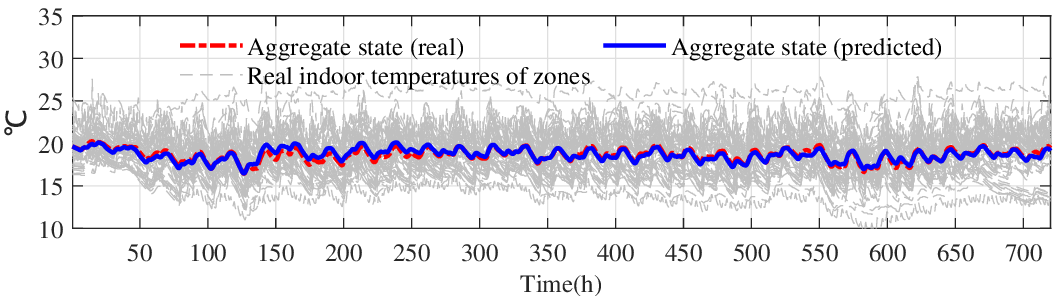}
 \setlength{\abovecaptionskip}{-5pt}
 \vspace{-0.2cm}
 \caption{Aggregate state and real indoor temperatures of zones.}
 \label{fig_result_A_2}
 \end{figure}

 Moreover, the real aggregate state and the predicted aggregate state of the building cluster under the privacy-preserved algorithm for ATDM are illustrated in Fig. \ref{fig_result_A_2}. The results show that our proposed algorithm has excellent accuracy in both the training and test sets.

\begin{table}[t]  
    \centering
    \footnotesize 
    \caption{Error Indicators}  
    \vspace{-0.2cm}
    \label{table_result_A_2}
    \begin{tabular*}{\linewidth}{c c c c}
    \toprule
    \makebox[0.15\linewidth][c]{\textbf{Indicator}} & 
    \makebox[0.17\linewidth][c]{\textbf{RMSE(℃)}} & 
    \makebox[0.17\linewidth][c]{\textbf{\textbf{MAPE (\%)}}} &
    \makebox[0.15\linewidth][c]{{$\textbf{R}^\textbf{2}$}}  \\
    \midrule
    \textbf{Privacy-preserved}     & 0.2944            & 1.3103             & 0.8613   \\
    \textbf{Non-privacy-preserved} & 0.2741            & 1.2127             & 0.8797   \\
\bottomrule
\end{tabular*}
\end{table}

\subsection{Masking Performance Analysis}
\begin{figure}[t]   
\captionsetup{singlelinecheck=on}
  \centering
  \footnotesize 
  \subfloat[]  
  {
      \label{fig_result_B_11}\includegraphics[width=1\linewidth]{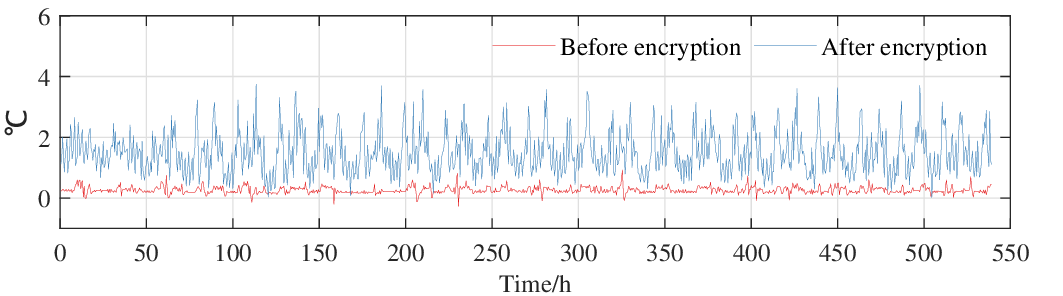}
  }\\
  \vspace{-0.2cm}
  \subfloat[] 
  {
      \label{fig_result_B_12}\includegraphics[width=1\linewidth]{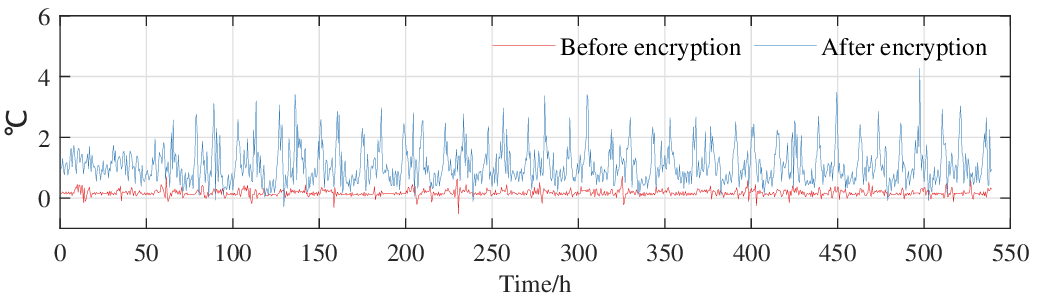}
  }\\   
    \vspace{-0.4cm} 
   \subfloat[] 
  {
      \label{fig_result_B_13}\includegraphics[width=1\linewidth]{figure/result_B_iter2.eps}
  }\\
  \caption{The first column of $\hat{\tau}_{in,z}$ and of sites encryption version $\hat{\tau}_{in,z} W^T$: (a) Iteration 1; (b) Iteration 2; (c) Iteration 3.}     
    \label{fig_result_B_1} 
\end{figure}

To measure the performance of masking information, we compare the original data and masked data. We take the first column of the original $\hat{\tau}_{in,z}$ and its encryption version $\hat{\tau}_{in,z} W^T$ in all three iterations as an example as shown in Fig. \ref{fig_result_B_1}. It is evident that the information, $\hat{\tau}_{in,z}$, is perfectly masked by the encryption matrix, $W^T$.


\subsection{Computational Performance Analysis}
Comparing the computing time under the privacy-preserved algorithm and the non-privacy-preserved algorithm for ATDM, problems of different scales are considered for comprehensive analysis. As shown in Table \ref{table_result_C_1}, when the computation scale increases, the computation time increases regardless of whether privacy preservation is considered. In addition, the computing time is significantly longer when privacy preservation is considered. This is because the generation and operation of large numbers of random matrices take abundant time when the SAP and TE methods are implemented. This shows a trade-off between the privacy preservation effect and the computing efficiency.

\begin{table}[t]
    \centering
    \footnotesize
    \caption{Computation Time and Iterations}
    \label{table_result_C_1}
    \vspace{-0.2cm}
    \begin{tabular}{cccccc}
    \toprule
    \multirow{2}{0.15\linewidth}{\textbf{No. of buildings}} & \multirow{2}{0.12\linewidth}{\textbf{Num. of agents}} & \multicolumn{2}{c}{\textbf{Privacy-preser.}} & \multicolumn{2}{c}{\textbf{Non-privacy-preser.}} \\
     \cmidrule(lr){3-6}
            & & Time (s)         & Iter.         & Time (s)          & Iter.         \\
    \midrule
     \textbf{\scriptsize 1}          & 10  & 0.2920         & 2        & 0.2763        & 2        \\
     \textbf{\scriptsize 1/8}         & 19   &  0.6121         & 3          & 0.3784          & 3          \\
     \textbf{\scriptsize 1/8/10}       & 27    & 1.0593          &   3        & 0.3865          &  3         \\
     \textbf{\scriptsize 1/8/10/11}      & 37    &  2.100          &    3       & 0.4117          &  3         \\
     \textbf{\scriptsize 1/8/10/11/13}     & 47      & 3.7237          &   3        & 0.4361          &  3         \\
     \textbf{\scriptsize 1/8/10/11/13/16}     & 56     &  5.6593         &  3         &  0.4672         &  3         \\
     \textbf{\scriptsize 1/8/10/11/13/16/20}    & 64       &  9.2325         & 3          & 0.4785          &  3         \\
         \bottomrule
\end{tabular}
\end{table}

\subsection{Privacy Security Analysis} 
 \label{section_V_D}

In this part, we analyze the privacy inference problem numerically. The "fmincon" solver in Matlab is used to obtain the least squares solution of the MQS problem \eqref{MQS}. We generate the component of $W^{(l)} (l\in \mathbf{L})$ and $\tau_{in,z}^{-m}$ using a normal distribution with a standard deviation of 1 and a mean of 20. Based on this, the constant matrices and vectors in \eqref{MQS} can be obtained. Then, the BLA can solve \eqref{MQS} and get the solution $\tau_{in,z,0}^{-m}$ and $W^{(l)}_0$. We can compare the differences between $\tau_{in,z}^{-m}$ and $\tau_{in,z,0}^{-m}$ to verify whether the BLA makes an accurate privacy inference. In our simulations, we set $K=6, L=3, T\in\{1,2,3,4,6,12,24,48\}$, 8 cases in total. The initial point is critical in solving the MQS \eqref{MQS}. We set 20 scenarios for each case, in which the initial point is set to the accurate value of $W^{(l)} (l\in \mathbf{L})$ plus random perturbations following a uniform distribution in $[-1, -1]$. 

\begin{figure}[t]  
 \captionsetup{singlelinecheck=on}
  \footnotesize 
  \centering     
  \subfloat[] 
  {
      
      \includegraphics[width=0.48\linewidth]{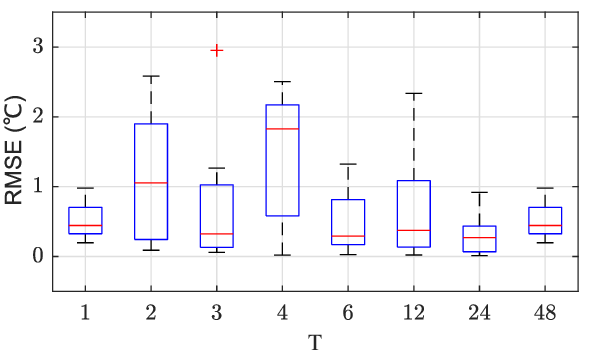}
      \label{fig_result_D_inference_errors}
  }   
  \subfloat[]  
  {
      \label{fig_result_D_solution_time}
      \includegraphics[width=0.48\linewidth]{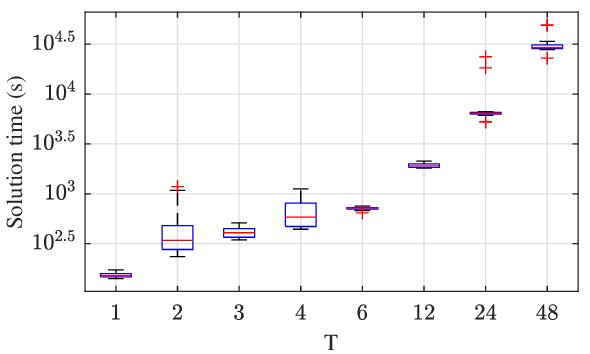}
  }
  \captionsetup{singlelinecheck=off}
  \caption{Privacy inference results: (a) Inference errors; (b) Computation time.}
  \label{fig_result_D}  
\end{figure}

The inference errors are given in Fig. \ref{fig_result_D_inference_errors}. The solution time of each case is shown in Fig. \ref{fig_result_D_solution_time}. Obviously, as the size of the MQS increases, the solution time increases exponentially. In practice, both $T$ and $K$ are significant numbers. Hence, we can conclude that obtaining the least squares solution of this MQS problem is impossible within an acceptable time. 

In summary, the simulation results reveal that (i) the inaccurate initial point leads to huge inference errors and (ii) the computational cost of the MQS increases exponentially. More importantly, even if BLA coincidentally infers the correct values, it cannot determine whether it is true because determining whether a local optimum is a global optimum is still an NP-hard problem for a nonconvex problem. Considering these points, the BLA cannot infer private information by solving the MQS problem \eqref{MQS}.

\section{Conclusion}
This paper proposes a privacy-preserved aggregate thermal dynamic model for the building cluster to provide an interface for the interaction between the energy system and buildings. The proposed privacy-preserved computation algorithm can estimate the parameters of the aggregate thermal dynamic model without exposing the privacy information of users. Simulation results based on real-world data demonstrate the excellent performance of the proposed method.

\appendices
\numberwithin{equation}{section}
\section{Privacy Disclosure Caused by $\hat{\tau}_{in,z}$} \label{appendix_a}

This section illustrates why the indoor temperatures will be disclosed if $\hat{\tau}_{in,z}^{(l)}$ is known to the BLA. Based on the original definition of $\hat{\tau}_{in,z}$ in \eqref{hat_tau_definition}, the BLA can have the inference equations about building zone $i\ (\forall i\in \mathbf{K})$ as
\begin{equation}
    \label{app_a_eq2}
    \hat{\tau}_{in,z}^{i,t,(l)}=\tau_{in,z}^{i,t}-\sum_{m=1}^{M}\alpha_{m}^{(l)}\tau_{in,z}^{i,t-m},\quad t\in \mathbf{T},l\in \mathbf{L},
\end{equation}
wherein there are $(T+M)$ unknown variables, i.e. $\tau_{in,z}^{i,1-M},\cdots,\tau_{in,z}^{i,T}$. On the other hand, there are $TL$ linear equations during all iterations. Considering the model order $M$ is usually a small integer like 2 or 3, the condition $TL>T+M$ will hold as long as $L\geq2$. Thus, the BLA can infer $\hat{\tau}_{in,z}^{i,t}\ (\forall i \in \mathbf{K}, \forall t\in \mathbf{T})$ when $L\geq2$. (The condition $L>2$ is easy to satisfy in practical situations.)  

\vspace{-0.2cm}
\section{Privacy Disclosure Caused by $W_i$} \label{appendix_b}
This section proves that $W^{[i]} (W^{[i]})^T$ will disclose information of the random matrix $w_{i,j}$.
\begin{equation}
    \label{app_b_eq1}  
    W^{[i]}(W^{[i]})^T=\begin{bmatrix}
    w_{1i}w_{1i}&\cdots&w_{1i}w_{Ki}\\
    \vdots&\ddots&\vdots\\
    w_{Ki}w_{1i}&\cdots&w_{Ki}w_{Ki}\end{bmatrix} .
\end{equation}

Once $W^{[i]}(W^{[i]})^T\ (\forall i \in \mathbf{K})$ is known to the BLA, it can easily get the proportional relationship between $w_{1i}$,$w_{2i}$,$\cdots$,$w_{Ki}$ based on the inference equation in \eqref{app_b_eq1}. 
\begin{equation}
    \label{app_b_eq2}
    w_{i1}=g_1 w_{i2}=g_2 w_{i3}=\cdots=g_{K-1} w_{iK},\ \forall i\in \mathbf{K},
\end{equation}
wherein $g_1$,$g_2$,$\cdots$,$g_{K-1}$ are known constants for the BLA. Moreover, during each iteration $l$, the BLA can acquire $\bar{\xi}$ calculated by itself and the recovered $\xi$ transferred from building zones. Thus, the BLA can infer $W^{[i]}\ (\forall i\in \mathbf{K}$) by jointly solving the equations \eqref{transformation_based_encryption_xi} and \eqref{app_b_eq2}.

\newcommand{\BIBdecl}{\setlength{\itemsep}{0.01 em}}
\bibliographystyle{IEEEtran}
\bibliography{IEEEabrv,References.bib}

\end{document}